\documentclass{appolb}
\usepackage{graphicx}
\usepackage{lineno}

\begin{document}
\title{Towards a Total Cross Section Measurement\\with the ALFA Detector at ATLAS%
\thanks{Presented at the Cracow Epiphany 2013 Conference.}%
}
\author{Maciej Trzebi\'nski\\
on behalf of the ATLAS Collaboration
\address{Institute of Nuclear Physics Polish Academy of Sciences\\152 Radzikowskiego St., Krak\'ow, Poland}\\
}
\maketitle
\begin{abstract}
The main goals of the Absolute Luminosity For ATLAS (ALFA) detector is to provide an absolute luminosity and total cross section measurement. The measurement method used, the detector alignment and the quality of the collected data are discussed.
\end{abstract}

  
\section{Introduction}
The main goal of the ALFA detector is to deliver information on absolute luminosity via the measurement of the elastic proton-proton scattering in the Coulomb-nuclei interference region \cite{ALFA_TDR}. Since in this region the four-momentum transfer is small (the Coulomb interaction starts to predominate for $|t| \sim 0.00065$ GeV$^2$) the protons are scattered at very small angles, typically of the order of several micro radians. Measurements in such an extreme kinematic domain require a special setting of the LHC optics as well as installations of the detectors far (typically hundreds of meters) from the Interaction Point (IP) and as close to the beam as possible.

It is worth noticing that, in addition to the elastic measurement, ALFA allows the detection of protons scattered diffractivelly. 

\section{The Idea of the Absolute Luminosity and the Total Cross Section Measurements}
The measurement of the absolute luminosity is based on the fact that the rate of elastic interactions is linked to the rate of all interactions through the optical theorem. This theorem states that the total cross section is directly proportional to the imaginary part of the forward elastic scattering amplitude at zero momentum transfer:
\begin{equation}
\sigma_{tot} = 4 \pi \cdot Im \left( f_{el}|_{t=0}\right).
\end{equation}
If $p$ is the momentum of scattered protons, then for small values of the scattering angle ($\theta$) one can write:
\begin{equation}
-t = (p\theta)^2.
\label{eq_t_angle}
\end{equation}
The rate of elastic scattering events at small $t$ values can be written as:
\begin{equation}
\frac{dN}{dt} = L \pi |f_C + f_N|^2 \approx 
L \pi \left|-\frac{2\alpha_{EM}}{|t|} + 
\frac{\sigma_{tot}}{4\pi}(i + \rho)\exp\left(\frac{-b|t|}{2}\right) \right|^2,
\label{eq_elastic_CS}
\end{equation}
where $f_C$ corresponds to the Coulomb and $f_N$ to the strong (nuclear) interaction amplitude, $b$ is the nuclear slope and $\rho$ is defined as:
\begin{equation}
\rho = \left. \frac{Re\ f_{el}}{Im\ f_{el}}  \right|_{t = 0}.
\end{equation}
From Equation \ref{eq_elastic_CS} one immediately notices that the $t$ measurement in the range where the nuclear amplitude dominates will give information about the total cross section, whereas at low $t$, where Coulomb scattering is dominating, the absolute luminosity can be precisely measured since the $f_C$ amplitude is well described by QED. One should also note, that Equation \ref{eq_elastic_CS} is simplified, since \textit{e.g.} the proton form factor was omitted\footnote{This factor is, however, included in the measurement.}.

The most suitable method to perform such a measurement is to use so-called ''parallel-to-point'' optics. In this type of optics there is a 90 degree phase advance between the Interaction Point and the detector position (only in the vertical plane in the solution used). This causes all particles scattered at the same angle to be focused on the same point on the detector, independently of the position of their interaction vertex. Thus, a transverse position measurement at the detector gives directly an angular measurement at the IP. Obviously, such optics requires a~special settings of the LHC magnets, as described in \cite{highbeta_optics}.

The LHC collides two proton beams which are circulating in two horizontally displaced beam pipes. The beam performing the clockwise motion (viewed from above the ring) is called $beam1$ and the other one $beam2$.

The $y$ position of a trajectory in the transverse plane at a given distance from the IP is given by:
\begin{equation}
\left[ 
\begin{array}{c} u(s) \\ u'(s) \\ (\Delta p)/p \\ \end{array}
\right] = M
\left[ 
\begin{array}{c} u^* \\ u'^* \\ (\Delta p^*)/p \\ \end{array}
\right],
\end{equation}
with the transport matrix
\begin{equation}
M =
\left[ 
\begin{array}{c c c} 
\sqrt{\beta/\beta^*}(\cos\psi + \alpha^*\sin\psi & \sqrt{\beta\beta^*}\sin\psi & D_u \\
\frac{(\alpha^* - \alpha)\cos\psi - (1 + \alpha\alpha^*)\sin\psi}{\sqrt{\beta\beta^*}} & \ \sqrt{\frac{\beta^*}{\beta}}(\cos\psi - \alpha\sin\psi) & D'_u \\
0 & 0 & 1 \\
\end{array}
\right],
\end{equation}
where $\beta$ is the betatron function\footnote{In terms of accelerator optics the betatron function $\beta$ is a measure of the distance from a certain point to the one at which the beam is twice as wide. The lower the betatron function at the IP ($\beta^{*}$), the smaller is the beam size, thus the larger is the luminosity.}, $\psi$ is the phase advance, $u$ denotes the transverse space coordinate ($x$ or $y$), $u'$ the corresponding trajectory slope, $D$ the dispersion acting on particles with a momentum loss $\Delta p/p$ and $\alpha$ is proportional to the derivative of the betatron function. Quantities at the Interaction Point are denoted by asterisk. Since for elastic scattering there is no energy loss and the dispersion acts only on the beam intrinsic energy smearing (which relative value is of the order of $10^{-4}$), the dispersion terms can be neglected. This implies the following formula of transport matrix for $y$:
\begin{equation}
\left(
  \begin{array}{c}
    y \\
    \theta
  \end{array}
\right) = 
\left(
  \begin{array}{c c}
    \sqrt{\frac{\beta}{\beta^*}}(\cos \psi + \alpha^* \sin \psi) & 
    L_y = \sqrt{\beta\beta^*}\sin \psi \\
    \frac{(\alpha^* - \alpha) \cos \psi - (1 + \alpha\alpha^*) \sin \psi}{\sqrt{\beta\beta^*}} &
    \sqrt{\frac{\beta^*}{\beta}}(\cos \psi  - \alpha \sin \psi)
  \end{array}
\right)
\left(
  \begin{array}{c}
    y^* \\
    \theta^*
  \end{array}
\right),
\end{equation}
from which the scattering angle: 
\begin{equation}
\theta^* = \frac{y^{beam1} - y^{beam2}}{L_{y}^{beam1} + L_{y}^{beam2}}
\end{equation}
and, thus, the $t$ value at the Interaction Point can be obtained (\textit{cf.} Eq. \ref{eq_t_angle}). Such way of extracting the $t$ value at the IP is called a 'subtraction method'.

\subsection{The ALFA Detector}
The details of the detector properties are given in \cite{ALFA_TDR} and here only a~few points are highlighted.

The ALFA experimental set-up consists of four detector stations placed symmetrically with respect to the ATLAS \cite{ATLAS} IP at 237.4 m and 241.5 m. In each station there are two Roman Pot devices. The stations on the A side (beam 2) are labelled A7L1 (first station; 237.4 m) and B7L1 (second; 241.5 m), whereas stations on the C side (beam 1) are labelled A7R1 (first station) and B7R1 (second station).

The main features of the ALFA detectors are: 
\begin{itemize}
  \item a spatial resolution of 30 $\mu$m \footnote{This value has to be considerably smaller than the spot size of the beam ($130\ \mu$m for $\beta^* = 2625$ m) at the detector location.}, 
  \item no significant inactive edge region, to be as close to the beam as possible,
  \item minimal sensitivity to the radio-frequency noise from the LHC beams.
\end{itemize}
The ALFA detectors were built using the scintillating fibre tracker technology. Since the detectors are used only during special LHC runs, when the instantaneous luminosity is low, the technology used is not radiation hard.

\subsection{The LHC Runs}
\begin{figure}[htb]
\centerline{%
\includegraphics[width=12.5cm]{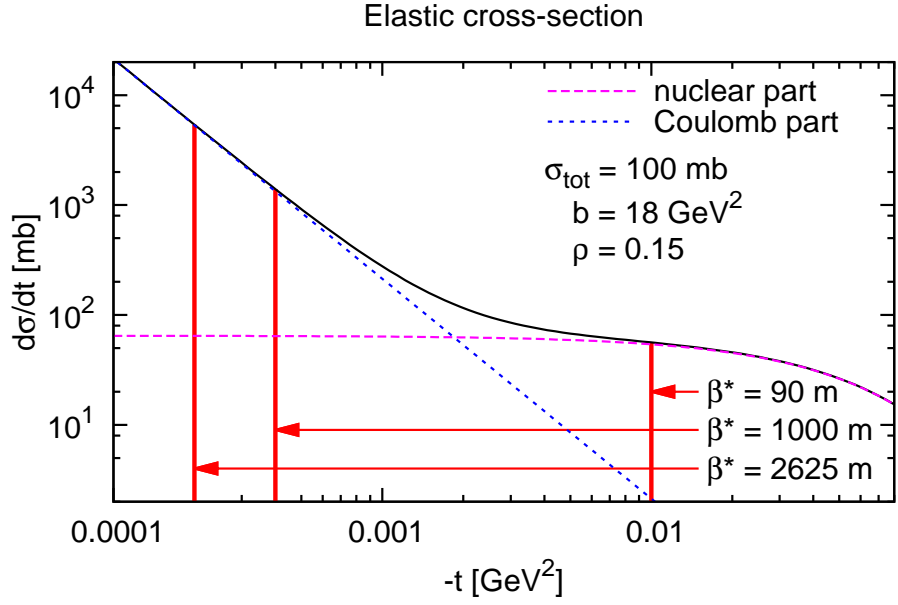}}
\caption{The elastic cross section as a function of $t$ for a possible set of parameters at LHC energies. An estimate of the expected $t$ value reachable at a given $\beta^*$ is plotted.}
\label{fig_t_range}
\end{figure}

During the LHC operation there were several proton-proton runs dedicated to special data taking, called \textit{high}-$\beta^{*}$ runs\footnote{One should note, that the betatron function value during standard LHC runs is $\sim 0.6$~m.}, when the ALFA detectors were allowed to be inserted into the LHC beampipe. So far, except for a few test runs, there were three \textit{high}-$\beta^{*}$ runs when the ALFA detectors collected data:
\begin{itemize}
  \item $\beta^{*} = 90$ m, with the centre of mass energy $\sqrt{s} = 7$ TeV, 
  \item $\beta^{*} = 90$ m, $\sqrt{s} = 8$ TeV, 
  \item $\beta^{*} = 1000$ m, $\sqrt{s} = 8$ TeV.
\end{itemize}  
It is worth mentioning that the betatron function value foreseen for optimal ALFA data taking is at $\beta^{*} = 2625$ m. An example of the minimum $t$ value that should be reachable with a given optics is shown in Figure \ref{fig_t_range}.

\section{Towards the Total Cross Section Measurement}
Precise measurement of the total cross section is not simple. Firstly, there is a need to have a very good alignment of the ALFA detectors. It has been done using several methods, based on surveys or on data. The final obtained precision in case of the relative (between the ALFA stations) alignment is of about 10 microns in horizontal $x$ and 5 microns in vertical $y$ axis. The precision of the absolute (w.r.t. the LHC beams) alignment is 60 $\mu$m in $y$, whereas in $x$ axis the stations are perfectly aligned.

\begin{figure}[htb]
\centerline{%
\includegraphics[width=12.5cm]{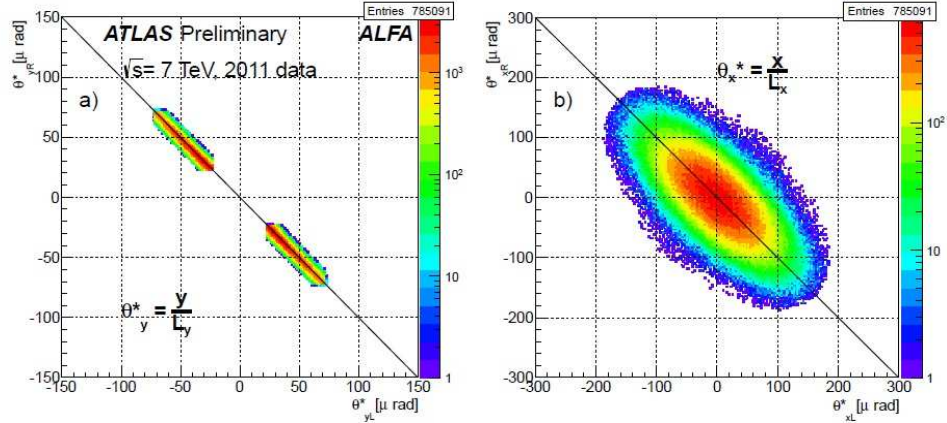}}
\caption{Reconstructed scattering angle correlation between A ($\theta^*_{xL},\ \theta^*_{yL}$) and C ($\theta^*_{xR},\ \theta^*_{yR}$) side for elastic candidates after background rejection cuts a) in the vertical and b) in the horizontal plane.}
\label{fig_ALFA_angle}
\end{figure}

The reconstructed scattering angle correlation between A and C side for elastic candidates a) in the vertical and b) in the horizontal plane after the selection is plotted in Figure \ref{fig_ALFA_angle} (from \cite{ATLAS_public}). The expected back-to-back topology is better visible in the vertical plane due to the LHC optics properties~\cite{trzebinski}.

The last step before having a precise measurement of the total cross section is to obtain a good knowledge of the transport matrix. This in turn requires the best possible knowledge of the LHC magnet settings between the ATLAS IP and the ALFA detectors. This task is recently under investigation by the ALFA analysis group.

\section{Summary}
The ALFA detector successfully took data during special LHC runs in years 2011 and 2012. The detectors are aligned with a local precision better than 10(5) $\mu$m in $x$($y$) and a global one of the order of 60 $\mu$m in $y$ axis. The behaviour of the data is well understood and the backgrounds are under control. The last missing point before obtaining the total cross section, is having a precise knowledge of the transport matrix elements.

As can be seen from Fig. \ref{fig_t_range}, runs with $\beta^* = 90$ m will allow for a total cross section measurement, whereas the one with $\beta^* = 1000$ m will probably allow to investigate the Coulomb-nuclear Interference region. The foreseen run with $\beta^* = 2625$ m should allow for a measurement in the region where the scattering amplitude is dominated by the QED, so that the precise determination of the absolute luminosity should be possible.

\end{document}